\documentclass[10pt,conference,letterpaper]{IEEEtran}
\usepackage{times,amsmath,epsfig}

\usepackage{subfigure}%
\usepackage{listings}%

\title{Enabling Lock-Free Concurrent Fine-Grain Access to Massive Distributed Data: Application to Supernovae Detection}

\author{%
{Bogdan Nicolae{\small $~^{\#1}$}, Gabriel Antoniu{\small $~^{*2}$}, Luc Boug\'e{\small $~^{+3}$} }%
\vspace{1.6mm}\\
\fontsize{10}{10}\selectfont\itshape
$^{\#}$\,University of Rennes~1/IRISA\\
Campus de Beaulieu, 35042 Rennes cedex, France\\
\fontsize{9}{9}\selectfont\ttfamily\upshape
$^{1}$\,Bogdan.Nicolae@inria.fr
\vspace{1.2mm}\\
\fontsize{10}{10}\selectfont\rmfamily\itshape
$^{*}$\,INRIA/IRISA\\
Campus de Beaulieu, 35042 Rennes cedex, France\\
\fontsize{9}{9}\selectfont\ttfamily\upshape
$^{2}$\,Contact: Gabriel.Antoniu@inria.fr\fontsize{9}{9}\selectfont\rmfamily\itshape
\vspace{1.2mm}\\
\fontsize{10}{10}\selectfont\rmfamily\itshape
$^{+}$\,ENS Cachan, Brittany Extension/IRISA\\
Campus Ker Lann, 35170 Bruz, France\\
\fontsize{9}{9}\selectfont\ttfamily\upshape
$^{3}$\,Luc.Bouge@bretagne.ens-cachan.fr%
}

\begin{document}

\maketitle

\begin{abstract}
  We consider the problem of efficiently managing massive data in a
  large-scale distributed environment. We consider data strings of
  size in the order of Terabytes, shared and accessed by concurrent
  clients. On each individual access, a segment of a string, of the
  order of Megabytes, is read or modified. Our goal is to provide the
  clients with efficient fine-grain access the data string as
  concurrently as possible, without locking the string itself. This
  issue is crucial in the context of applications in the field of
  astronomy, databases, data mining and multimedia.  We illustrate
  these requiremens with the case of an application for searching
  supernovae. Our solution relies on distributed, RAM-based data
  storage, while leveraging a DHT-based, parallel metadata management
  scheme. The proposed architecture and algorithms have been validated
  through a software prototype and evaluated in a cluster environment.
\end{abstract}

\section{Introduction}
\label{sec:intro}

Large scale data management is becoming increasingly important for a
wide range of applications, both scientific and industrial: modeling,
astronomy, biology, gouvernamental and industrial statistics, etc.
All these applications generate huge amounts of data that need to be
stored, processed and eventually archived globally. In order to better
illustrate these needs, this paper focuses on a real life astronomy
problem: finding supernovae (stellar explosions).

In a typical scenario, a telescope is used to take pictures of the
same part of space at regular intervals, usually every
month. Corresponding digital images are then compared in an attempt to
find variable objects, which might be candidates for supernovae. To
confirm that such objects are supernovae, considerable computational
effort is necessary in order to distinguish the supernovae themselves
from the other variable objects that may be present in the image: this
requires to analyze the light curve and spectrum of each potential
candidate.

To speed up the process of finding supernovae, multiple parts of space
should be analyzed concurrently: as there is no dependency between
different regions of space, the analysis itself is an embarrassingly
parallel problem. The difficulty lies in the massive amount of data
that needs to be managed and made available to the machines providing
the computational power.

\emph{Huge data size.} Hundreds of GB of images from various parts of
the sky may correspond to a single point in time. Since the analysis
requires multiple consecutive images of the same part of the sky, the
order of TB is quickly reached.

\emph{Global view.} Managing independent images manually is
cumbersome. Applications finding supernovae (and not only) are much
easier to design if a global view of the sky is available: finding the
right image at a given time simply translates into accessing the right
part of the sky view for that time.  Let us consider a very simple
abstraction of this problem, in which the view of the sky is a very
long string of bytes (blob), obtained by concatenating the images in
binary form.  Assuming all images have a fixed size, a specific part
of the sky is accessible by providing the corresponding offset in the
string. A simple transformation from two-dimensional to unidimensional
coordinates is sufficient.

\emph{Efficient fine grain access.} While many images make up the
global view of the sky, each of them needs to be accessed
individually. As each image is much smaller than the size of the
string representing the sky, fine-grain access to substrings is
crucial.

\emph{Versioning.} As new images are taken by the telescope, the view
of the sky needs to be updated, while the previous views of the sky
still need to be accessible.  It is desirable to refer to views of the
sky at particular moments in time, therefore versioning is necessary.

\emph{Read-read concurrency.} Comparison of images for different parts
of the sky is a massively parallel problem.  That is, concurrent reads
of different images in a view or concurrent reads of the same image in
different views should be efficiently processed in parallel.

\emph{Read-write concurrency.} The telescope may gather and store new
pictures (i.e. new versions of some part of the sky) while the
analysis proceeds on the previous versions. Consequently, in our
model, it is important to allow new versions of our global string to
be generated and written while the earlier versions are read and
analyzed: read-write concurrency is highly desirable for efficiency.

\emph{Write-write concurrency.} As multiple telescopes may be
available for gathering pictures from different parts of the sky, it
is also desirable for the storage system to efficiently support
concurrent writes: concurrent substring updates should generate the
new corresponding strings in parallel.

Our case study clearly illustrates typical requirements for the more
general problem of massive data analysis: storage of massive data,
efficient fine grain access to small data sets, snapshoting support.
These requirements need to be addressed in a space efficient way (by
sharing common parts of snapshots) and in a performance efficient way
(by supporting read/read, read/write and write/write
concurrency). Such requirements are also exhibited by many other types
of applications: databases (\cite{postgresql, conc_control,
  performance_models}), large-scale, continuous data mining
(\cite{data_mining}), multimedia (\cite{multimedia}), etc.

To address these requirements, one may rely on scalable distributed
file systems, which provide a familiar, file-oriented API
allowing to transparently access physically distributed data through
globally unique logical file paths. 
A very large distributed storage space is thus made available to
existing applications that usually use file storage, with no need for
modifications. This approach has been taken by a few projects like
GFS~\cite{GFS}, GFarm~\cite{gfarm}, GridNFS~\cite{gridnfs},
LegionFS~\cite{legionfs}, etc. Note however that most such approaches
are not highly optimized to efficiently support highly-parallel,
fine-grain accesses to the same file, especially when some concurrent
accesses modify the file. A similar, RAM-based approach is provided by
the concept of \emph{grid data-sharing
  service}~\cite{AntBerCarDesBouJanMonSen06GDS}, illustrated by the
JuxMem~\cite{AntBouJan05SCPE} platform. However, in JuxMem data blocks
are not not fragmented, so the largest data block that the service is
able to store is limited by the size of the RAM of a \emph{single}
node. 

In this work, we explore the possibility of simultaneously addressing
massive data storage, with efficient fine-grain access optimized for
high read-read, read-write and write-write concurrency. As opposed to
grid file systems, our service mainly relies on RAM storage. This
favors access efficiency, while data persistence can still be provided
following the scheme described in~\cite{AntCudGhaTat08Euro-Par}. Our
paper is organized as follows. Section~\ref{sec:specs} restates the
specification of the problem in a more formal
way. Section~\ref{sec:design} provides an overview of our algorithmic
design and precisely describes how data access operations are
handled. Concurrency issues are discussed in
Section~\ref{sec:concurrency}. Section~\ref{sec:impl} provides a few
implementation details and reports on a preliminary experimental
evaluation on a multi-site grid testbed. On-going and future work is
discussed in Section~\ref{sec:conclusion}.

\section{Specifications}
\label{sec:specs}

We focus on managing massive binary strings (in the order of TB) in a
highly concurrent environment. We introduce two further denominations
used throughout this paper: A \emph{page} is any substring whose size
is fixed (\emph{pagesize}) and whose offset is a multiple of
\emph{pagesize}. A \emph{segment} is any concatenation of consecutive
pages. By convention, both the \emph{size} of the strings we
manipulate and \emph{pagesize} are powers of 2. We define two
primitives to access strings:
\begin{center}
  \lstinline$vw = WRITE(id, buffer, offset, size)$
\end{center}
A \lstinline$WRITE$ results in \emph{patching} the string identified
by \emph{id} with the contents of the local \emph{buffer} of length
\emph{size} at the specified \emph{offset}. This generates a new
incremental snapshot of the string, identified by its version number:
the returned value \emph{vw}. Version numbers are successive integers
starting with~0, which is the initial version.  (By convention,
version 0 is the all-zero string.) The generated snapshot is the view
resulting from the successive application of all previous patches
(including the current one). At this point, the string version
\emph{rw} is said to be \emph{published}. We obviously want that each
\lstinline$WRITE$ will eventually publish its version (\emph{liveness}).
\begin{center}
\lstinline$vr = READ(id, v, buffer, offset, size)$
\end{center}
A \lstinline$READ$ results in filling \emph{buffer} with the segment
identified by \emph{(offset, size)} of string \emph{id}. This segment
is extracted from version \emph{v} if \emph{v} has already been
published. The returned value \emph{vr} is then the number of the
latest \emph{published} version of the string and $vr \ge v$ holds. If
\emph{v} has not yet been published, then the read fails.

Observe that the above conditions guarantee that all non-failing
\lstinline$READ$ operations on the same version \emph{v} and same
\emph{offset} and \emph{size} will yield the same substring.  This
substring is the segment \emph{(offset, size)} which is obtained by
successively applying the first \emph{v} patches to the initial
string. This ensures that all \lstinline$READ$ operations ``see'' the
\lstinline$WRITE$ operations in the same order. Everything happens as
if the patches had been applied in the same successive order. This is
a variant of \emph{global serializability}.

For completeness, we provide an additional primitive allowing to
allocate storage space (\lstinline$ALLOC$), which generates a
globally unique \emph{id}.

\section{Design}
\label{sec:design}

\begin{figure*}[t!]
     \centering
     \includegraphics[type=eps,ext=.eps,read=.eps,width=.90\textwidth]{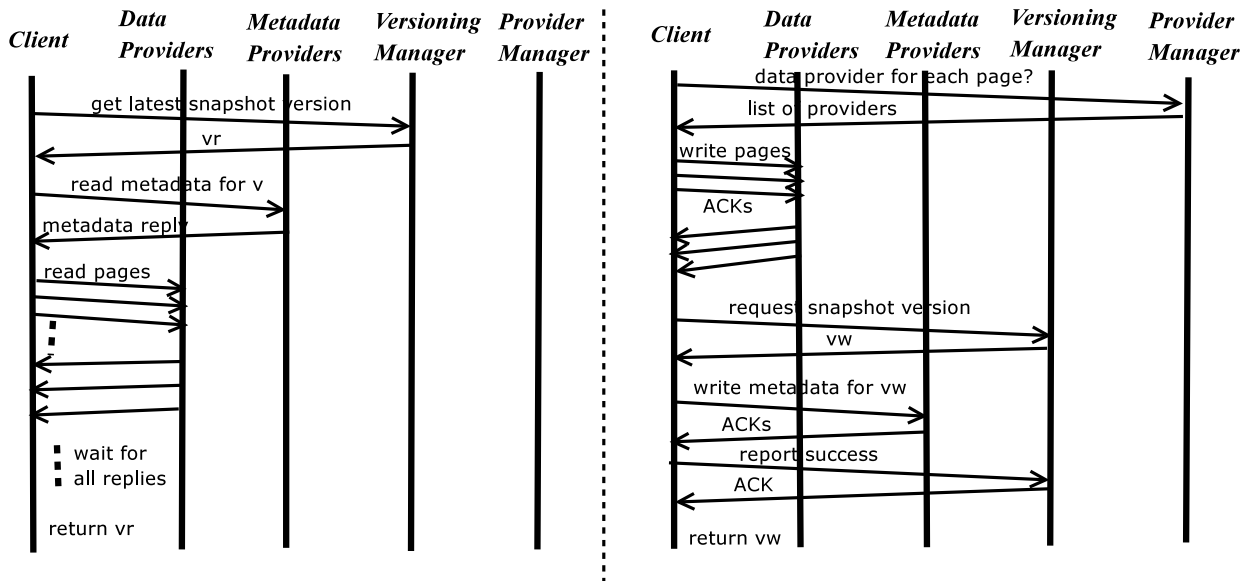}
     \caption{Interactions between the actors: reads(left) and writes(right)}
     \label{fig:interactions} 
\end{figure*}

Our system is striping-based: the set of \emph{pages} which make up
the global binary string is distributed among multiple
nodes. \emph{Metadata} defines the association between an access
request defined by \emph{(v, offset, size)} and the
corresponding set of pages storing the actual data. A
\lstinline$WRITE$ operation generates a new list of \emph{fresh} pages
stored on potentially new physical nodes. This way, no page is deleted
from the system at that time: the previous version of the pages remain
available through \lstinline$READ$ requests until some garbage
collection is ordered by the client. Each page is labeled with the
corresponding version number.

\subsection{General architecture overview}
\label{sec:arch}

Five kinds of actors make up the system:

\emph{Clients} issue \lstinline$READ$ and \lstinline$WRITE$
requests. There may be multiple concurrent clients. Their number may
dynamically vary in time without notifying the system.

\emph{Data providers} physically store in their local memory the pages
created by the \lstinline$WRITE$ operations. New data providers may
dynamically join the system.

\emph{A provider manager} keeps a information about the available data
providers.  On entering the system, each data provider register with
the provider manager. On each \lstinline$WRITE$ request, the provider
manager decides which providers should be used to store the newly
generated pages, based on some strategy that favors global load
balancing.

\emph{The metadata provider} physically stores the metadata allowing
generated when new pages are created by \lstinline$WRITE$ requests.
This entity is queries by clients issuing \lstinline$READ$ requests,
in order to find the pages corresponding to the requested range and
version.  Note that this can be a distributed entity, based on an
off-the-shelf distributed hash table (DHT), which allows efficient
concurrent access to metadata.
 
\emph{The version manager} is the key actor of the system. It stores
the number of the latest published version of a given data string. It
is also responsible for serializing \lstinline$WRITE$ requests to each
string, and for supplying \lstinline$READ$ requests with the latest
published string version. 

Our service consists of distributed communicating processes. Their
interaction is described below. In a typical setting, each process
runs on a separate physical node. A node may fulfill a specific role
by running a single process, but it may also play multiple roles.

\subsection{How reads and writes work}
\label{sec:how}

The interactions between the entities of our architecture are briefly
illustrated on Figure~\ref{fig:interactions}, both for
\lstinline$READ$ (left) and \lstinline$WRITE$ requests (right). For a
\lstinline$READ$ request, the client contacts the version manager to
get the latest version available for the corresponding string. If the
specified version is available the client contacts the metadata
provider to retrieve the metadata describing the pages of the
requested segment at the requested version. This operation results in
sending and processing parallel requests to the metadata providers
(as metadata are distributed and stored on a DHT). Once the client
gathers all the metadata, it contacts (in parallel again) the data
providers that store the corresponding pages and downloads them into
the local buffer.

On issuing a \lstinline$WRITE$ request, a client first contacts the
provider manager to get a list of providers, one for each page of the
segment to be written. The client then contacts (in parallel) the
corresponding providers and requests them to store the respective
pages. Each provider executes the request and sends an acknowledgment
to the client. When the client has received all acknowledgments, the
client contacts the version manager and requests a new version
number. This version number is then used by the client to generate the
corresponding new metadata. Then the client sends these metadata to
the metadata provider (in parallel again) and waits for an
acknowledgment.  Finally, the client contacts the version manager and
reports success.

Note that both for \lstinline$READ$ and \lstinline$WRITE$ requests,
the only serialization occurs when interacting with the version
manager. These interactions are reduced to simply requiring a version
number: all the other steps are fully parallel.

\subsection{Metadata management}
\label{sec:metadata}

Metadata store information about the pages which make up a given data
string, for each version available in the system.  Our goal is to
support fast metadata query for the \lstinline$READ$ requests, fast
metadata update for the \lstinline$WRITE$ requests, and to minimize
the overall metadata storage space in the system.

\begin{figure*}
  \centerline{%
  \hfill
  \subfigure[A segment tree: each node covers
  \emph{(offset, size)}, leaves refer to the pages]%
  {\label{fig:metadata_organization}%
    \includegraphics[type=eps,ext=.eps,read=.eps,width=.22\textwidth]
    {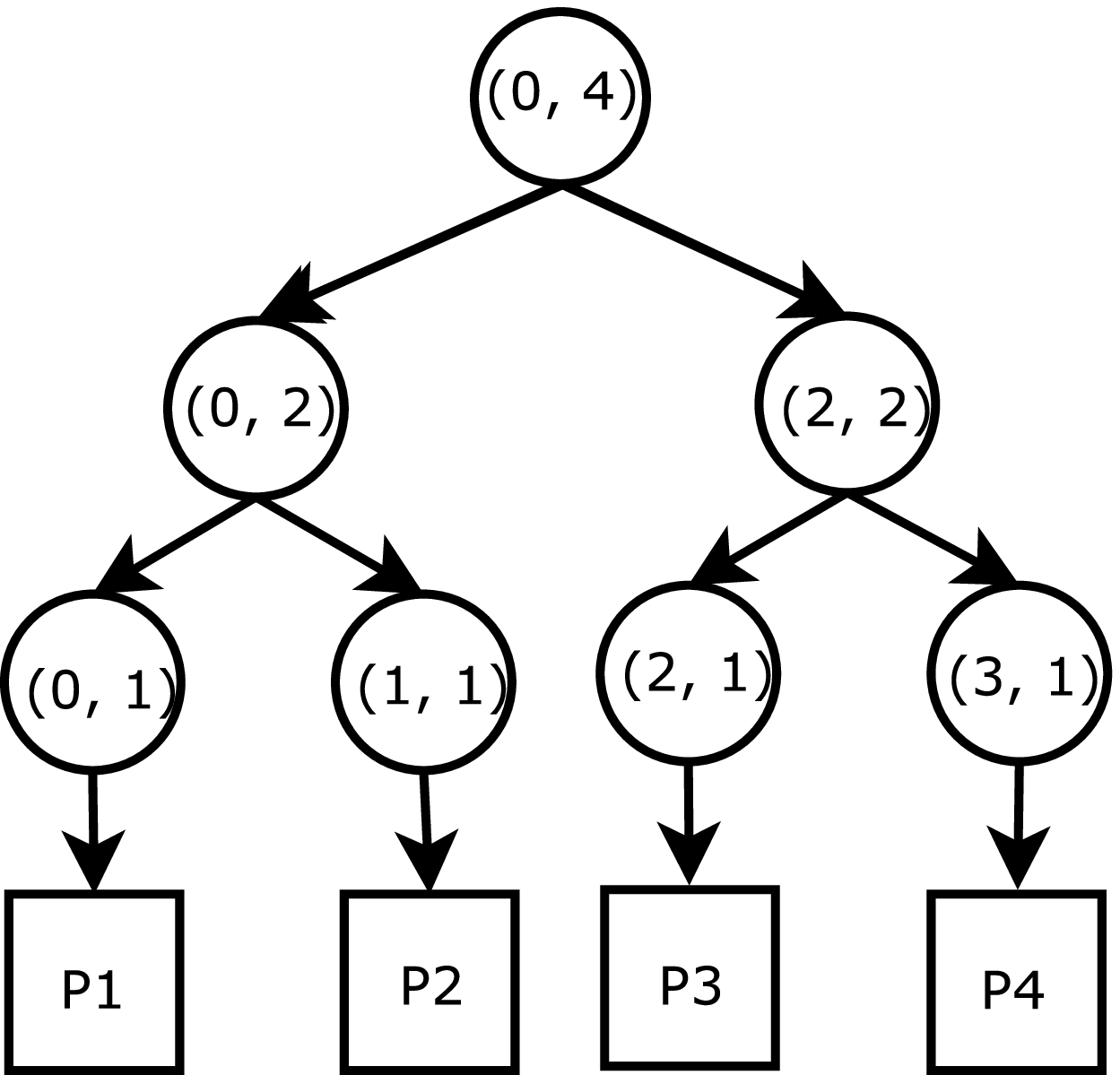}}
  \hfill
  \subfigure[Constructing new metadata: colored
  nodes are generated and linked to the previous version]%
  {\label{fig:metadata_rebuild}
    \includegraphics[type=eps,ext=.eps,read=.eps,width=.22\textwidth]%
    {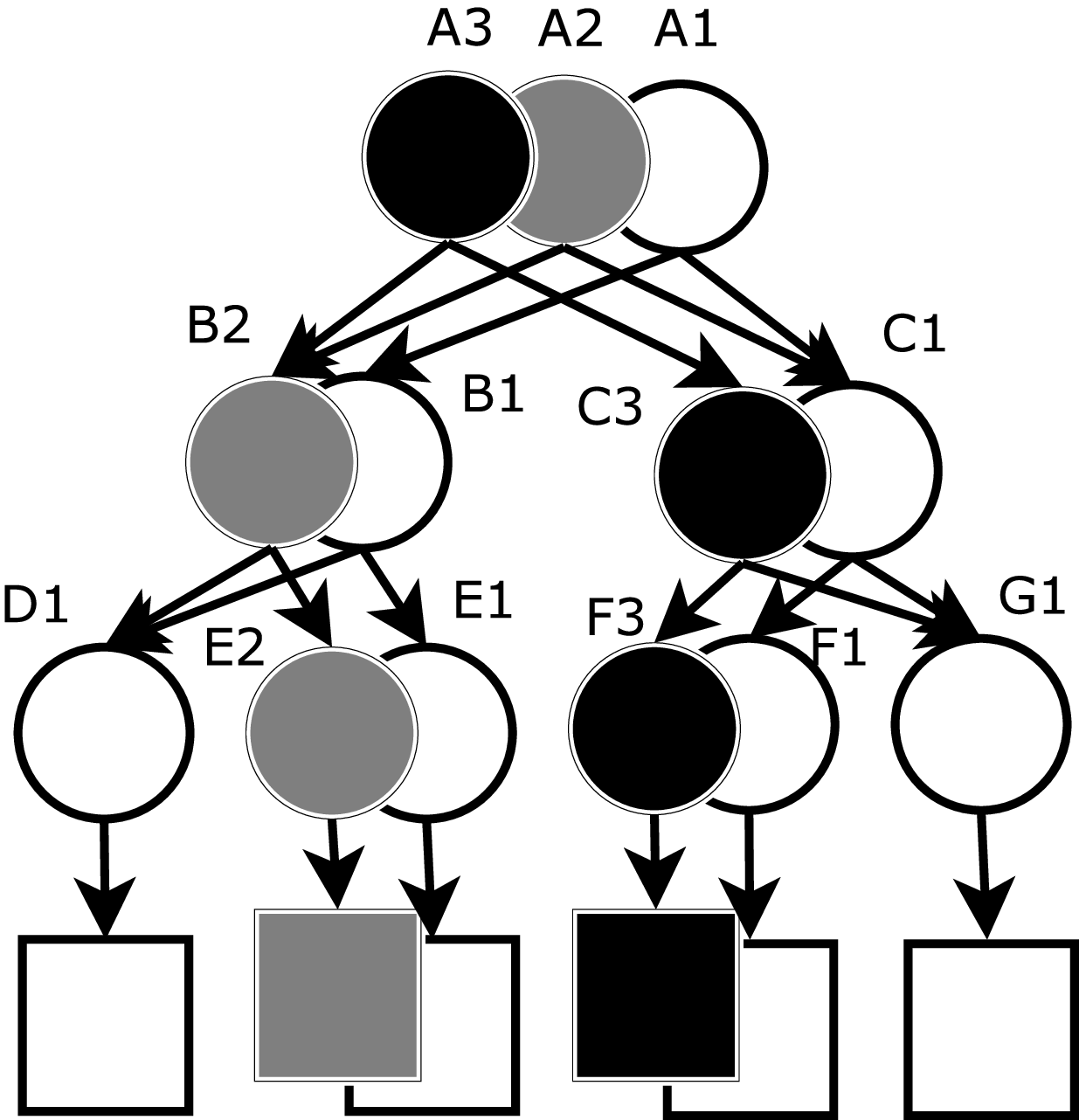}}
  \hfill
  }
  \caption{Metadata representation of a 4-page block}%
  \label{fig:Reads and writes}%
\end{figure*}

We organize metadata as a distributed \emph{segment
  tree}~\cite{segment_tree}, one associated to each version of a given
string \emph{id}. It is a full binary tree, with each node of
associated to a segment in the string identified by \emph{offset} and
\emph{size}. Such a node is said to \emph{cover} the segment. The left
child of the node covers the first half of the segment and the right
child the other half, with leaves covering a single page.  The node
stores additional information: the global string \emph{id} and its
version number \emph{v} (Figure~\ref{fig:metadata_organization}). To
find the pages making up a segment, one must traverse down the segment
tree, starting from the root. A node is visited only if its covered
interval intersects the segment. All leaves reached this way
correspond to the pages that are part of that segment. For example, in
Figure~\ref{fig:metadata_organization}, the set of nodes explored for
segment $[1, 2]$ is ${(0, 4), (0, 2), (2, 2), (1, 1), (2, 1)}$.  Out
of these, $(1, 1)$ and $(2, 1)$ are the leaves and refer to the
pages of segment $[1, 2]$.

A \lstinline$WRITE$ request producing version \emph{v} of a given
string needs to build a new metadata tree. This tree is the smallest
(possibly incomplete) binary tree of the same height as the initial
tree such that its leaves are exactly the leaves covering the pages of
the patched segment.  The leaves of this new tree exactly refer to
these pages that are part of the segment. This incomplete metadata
tree needs to be ``weaved'' into the previous complete metadata tree
such that its incomplete nodes (having a single left or a single right
child and referred to as \emph{border nodes}) will become complete by
referring to the missing corresponding child in the metadata tree
corresponding to the previous
version. Figure~\ref{fig:metadata_rebuild} illustrates this feature
through a simple scenario, in which the the initial version (white) is
1. The \lstinline$WRITE$ request on segment $[1, 1]$ is assigned
version 2 (grey). Its tree, is woven into the white tree: the missing
left child of $B_2$ is set to $D_1$ and the missing right child of
$A_2$ is set to $C_1$.  Similarly, a consequent \lstinline$WRITE$
request on segment $[2, 1]$ is assigned version 3
(black). Interweaving with the previous tree (gray) translates into
setting the right child of $C_3$ to $G_1$ and the left child of $A_3$
to $B_2$. Once they are built, the metadata tree nodes are uniformly
dispersed among the metadata providers (through the underlying DHT).


%

\section{Managing concurrency}
\label{sec:concurrency}

\subsection{Enabling parallel reads}

Dealing with concurrent reads is straightforward. As explained in
Section~\ref{sec:design}, each client starts by requesting the latest
version available from the version manager. This step (whose cost is
negligible with respect to the following steps) is the only
interaction with a centralized entity. Then each traverses the tree
down to the leaves to fetch the corresponding pages. Both tree
traversal page fetching can be performed by clients with full
parallelism, with no synchronization necessary with respect to other
clients. This is favored by the fine-grain dispersal of both data and
metadata across the distributed nodes.

\subsection{Enabling parallel reads with respect to concurrent writes}

As explained above, reads are performed in total isolation by each
client once the latest version is received from the version
manager. The only possible conflict with a concurrent write request
may occur at the level of the version manager, when a writer
increments the latest published version number. Consequently, the
relative cost due to such a potential conflict is negligible with
respect to the total access cost, we may consider that accesses are
fully parallel.

\subsection{Enabling parallel writes}

As explained in Section~\ref{sec:design}, \lstinline$WRITE$ operations
involve two phases: writing the data (i.e., the pages), then writing
the metadata (i.e., creating the metadata tree nodes). For any
concurrent \lstinline$WRITE$ operations to segments of the same
string, pages may be written in parallel with no synchronization. This
holds even when the \lstinline$WRITE$ operations concern non-disjoint
segments of the string, as each written segment involves a new set of
pages to be stored on potentially new data providers. Remember that
data is never actually modified: the old version of the data still
remains available on some providers.

Building and writing new metadata tree nodes might seem to require
serialization. Writing a newer version implies weaving the metadata
subtree into the full metadata tree of the previous version, as
explained in Section~\ref{sec:design}.
Even when the previous version is being written concurrently, we can
actually predict the missing children for the border nodes at a slight
computation overhead on the side of the versioning manager, no matter
how many concurrent writes compete for metadata weaving. Due to space
constraints, we do not develop the details of this mechanism here. 
Getting a precomputed set of border nodes from the version manager
enables the writer to generate the metadata in complete isolation with
respect to the other writers. After metadata is written, the client
reports success to the version manager. 

%

\section{Experimental evaluation}
\label{sec:impl}

\begin{figure*}[t]
  \centerline{%
    \subfigure[Metadata overhead, single client: reads]%
    {\label{fig:meta_read_same}
      \includegraphics[width=0.33\textwidth]{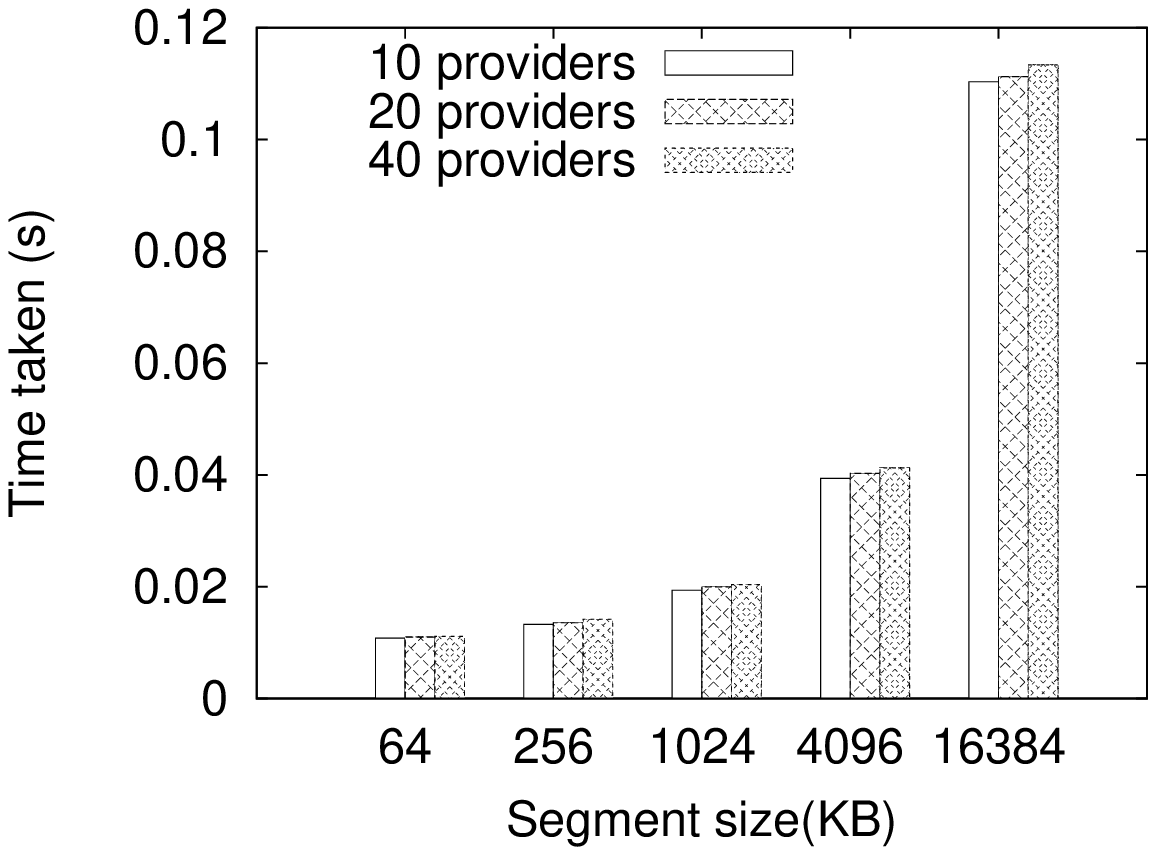}}
    \subfigure[Metadata overhead, single client: writes]%
    {\label{fig:meta_write_same}
      \includegraphics[width=0.33\textwidth]{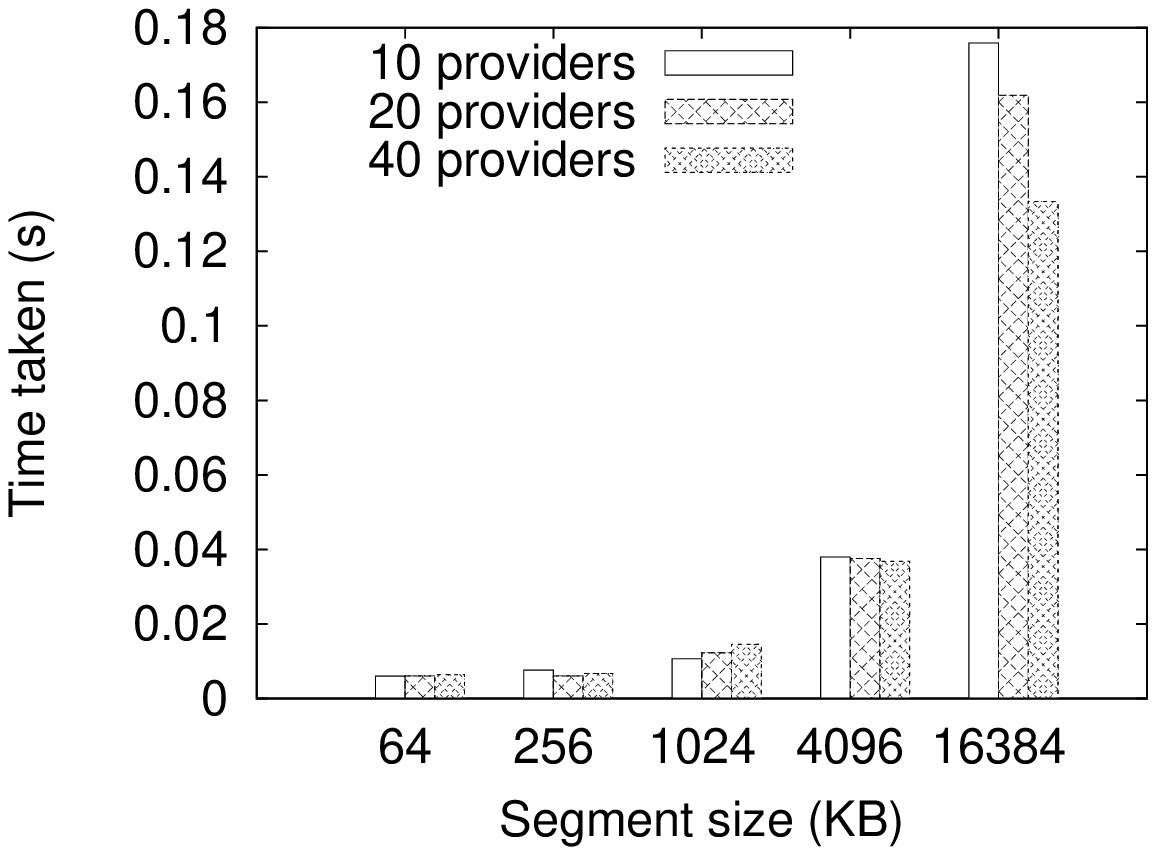}}
    \subfigure[Throughput of concurrent client access]
        {\label{fig:multi_same}
          \includegraphics[width=0.33\textwidth]{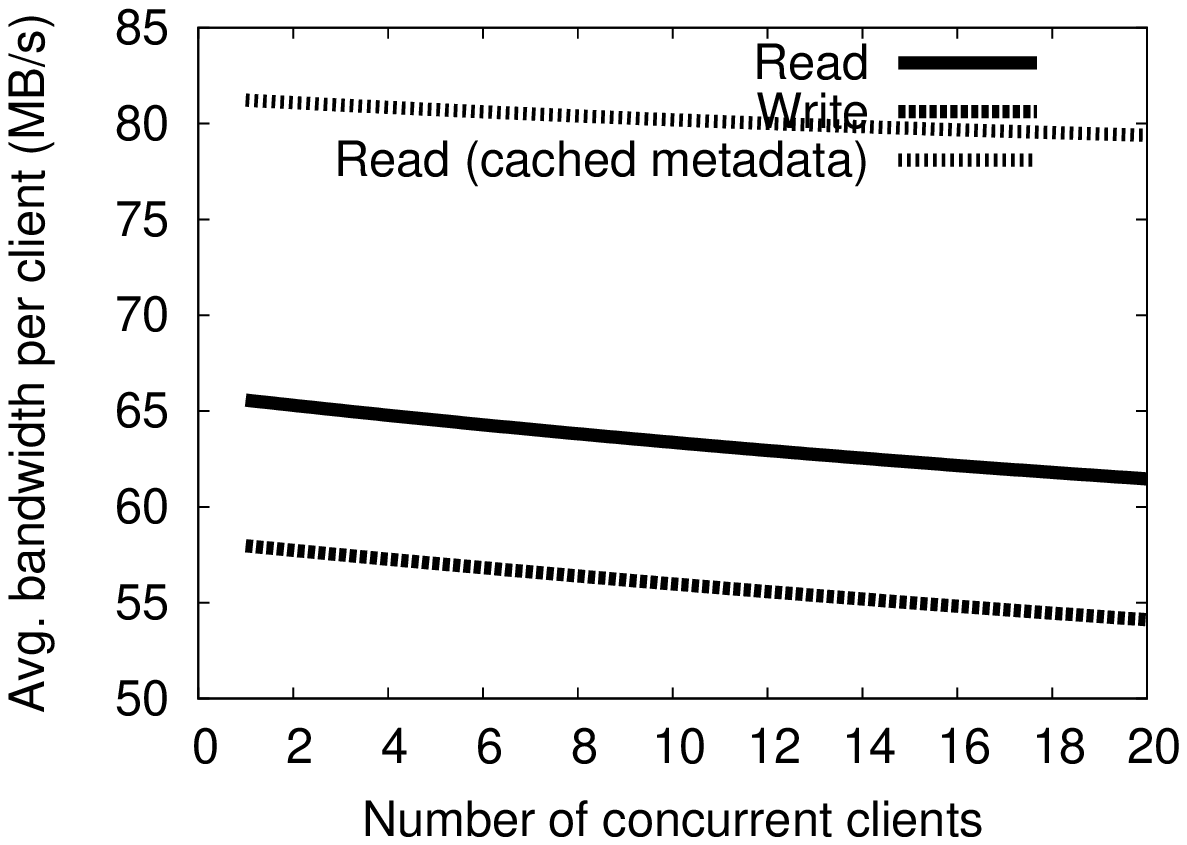}
        }
  }
  \caption{Metadata overhead for a single client and throughput for concurrent clients when nodes are in the same cluster (latency = 0.1~ms)}
  \label{fig:results_same}
\end{figure*}


\subsection{Implementation details}

Our implementation is based on the Boost C++ collection of
libraries~\cite{boost}.  We chose Boost for its standardization
throughout the C++ community, and for the wide range of
functionalities it provides, among which serialization, threading and
asynchronous I/O are of particular interest to us. For metadata
storage and retrieval, we use BambooDHT \cite{bamboo}, a stable,
scalable DHT implementation on top of which we build the abstraction
of our metadata providers.

Processes in our system communicate through RPCs. We allow a single
client to perform a large number of concurrent RPCs to enhance parallelism
and turn fine grain dispersion of data and metadata in our advantage.
However there is a tradeoff between striping and streaming. Dispersing data 
too fine grained might not pay off because of RPC call overhead.
For this reason we use lightweight custom RPC framework, which delays 
RPC calls to a single machine and streams all of them in a singe real
RPC call.


\subsection{Experimental platform}

Evaluations have been performed using the Grid'5000~\cite{grid5000_sc05}
testbed, a reconfigurable, controllable and monitorable experimental
Grid platform gathering 9 sites geographically distributed in
France. We used 50 nodes from a cluster located on the Grid'5000 site
in Rennes. Nodes are outfitted with x86\_64 CPUs and 4~GB of RAM, and
run Ubuntu (Linux 2.6). Intracluster bandwidth is 1 Gbit/s (measured:
117.5MB/s for TCP sockets with MTU = 1500~B), latency is 0.1~ms.

\subsection{Metadata overhead}

As a major goal of our system is to allow applications to store huge
data (of the order of 1~TB), we first evaluate how our metadata scheme
impacts the performance of data accesses. We first consider a single
client which allocates 1~TB of memory, then accesses a segment varying
from 16~KB to 16~MB. Note that the system allocates on write, which
means that only the segments that are written are physicaly allocated.
The data and metadata are distributed among a varying number of data
providers and metadata providers.  We successively use 10, 20 and 40
distinct physical nodes, each hosting one data provider and one
metadata provider.  The provider manager and the version manager are
deployed on separate, dedicated nodes.


We measure the time it takes for metadata to be completely read
(respectively written) for a \lstinline$READ$ (respectively
\lstinline$WRITE$), for a 1~TB string, using 64~KB
pages (Figure~\ref{fig:results_same}).


We observe that increasing the number of providers has a small impact
on the cost perceived by the client issuing a  \lstinline$READ$ request. For a
fixed number of tree nodes distributed on a variable number of
metadata providers, the retrieval cost perceived by the client is
almost the same. In fact, using a larger number of metadata providers
slightly increases the overall cost, as the client needs to manage
more connections. The main limiting factor is actually the performance
of the client's processing power. However, there is a benefit in using
a large number of metadata providers: this improves the reactivity of
the metadata providers when they are under heavy load, in conditions
of high access concurrency, because of the better load balancing.

In the case of \lstinline$WRITE$ requests, our observation is
different: using a larger number of metadata providers improves the
cost of writing the overall metadata. This is explained by our
optimized RPC mechanism, which aggregates requests for storage sent to
the same remote process. This is more visible when writing larger
segments.

\subsection{Throughput of concurrent clients}

Our second experiment aims at evaluating the efficiency of our 
lock-free scheme in a highly-concurrent environment. We measure the
average bandwidth per client for \lstinline$READ$ (respectively
\lstinline$WRITE$) requests when increasing the number of simultaneous
readers (respectively writers). We use 20 distinct nodes to deploy 20 reader
clients and another 20 physical nodes, each of which hosts one data
provider and one metadata provider. The version manager and the
provider manager run on another two dedicated physical nodes. The same
configuration is used with writers instead of readers. 

The experiments run as follows. First, a data string of 1~TB is
allocated, using tiny, 64-KB pages (in order to generate a access
various disjoint segments within a 1~GB interval of the data string in
a 100-iteration loop. Clients start simultaneously, then run without
any synchronization. As illustrated on
Figure~\ref{fig:results_same}, in all settings, we can notice
that the per client bandwidth hardly decreases when the number of
concurrent clients significantly increases. Besides, note that this
read bandwidth corresponds to a worst-case experiment, in which
client-level caching has been totally disabled! Read bandwidth is much
higher in real life situations, where client-side caching of metadata
tree nodes results in optimizing out a large amount of RPC calls. In
our experiments, the cache can accommodate $2^{20}$ tree nodes.

\section{Conclusion}
\label{sec:conclusion}
\enlargethispage{-15mm}

We address the problem of efficiently managing massive data in a
distributed environment. As a case study, we consider a problem in the
field of astronomy, consisting in searching for supernovae in a huge
set of images representing the sky at various moments in time.  Our
problem illustrates typical requirements for massive data analysis:
storage of massive data, efficient fine grain access to small data
sets, snapshoting support, with efficient read/read, read/write and
write/write access concurrency. We consider binary strings of size in
the order of Terabytes, which are intensively accessed by a set of
concurrent clients. For each such access, only a tiny segment of such a
string, of the order of Megabytes, is read or modified.

Our contribution is to propose an algorithm and system design which
let the clients access the strings as concurrently as possible,
without locking the string itself. Efficient fine-grain access to
arbitrarily tiny parts of the data is provided thanks to a distributed,
memory-based storage of individual pages, while leveraging a DHT-based,
inherently parallel metadata management scheme. 

Preliminary experiments have been run on a cluster from the Grid'5000
testbed. It turns out that our approach scales well, both in terms of
storage providers and in terms of concurrency degree: the per-client
bandwidth remains high when the number of concurrent clients
increases.

Our prototype is however a work in progress and needs further
refinement. First, fault tolerance, which becomes critical in
large-scale grid environments, is only partially addressed through the
use of the off-the-shelf DHT which implements the metadata
provider. We plan to also include fault-tolerance mechanisms for the
entities that currently represent single points of failure (version
manager, provider manager). Second, we also intend to address the
issue of garbage collection. Finally, we intend to realize large-scale
experimens with real applications in the fields of databases and data
mining.

{\small
\bibliographystyle{IEEEtran}
\bibliography{paper}
}

\end{document}